\documentclass{emulateapj}
\usepackage{graphicx}

\usepackage{natbib} 

\begin{document}

\title{Twenty-One New Light Curves of OGLE-TR-56b: New System Parameters and Limits on Timing Variations\altaffilmark{1}}

\author{E. R. Adams\altaffilmark{2,3}, M. L\'opez-Morales\altaffilmark{4,5}, J. L. Elliot\altaffilmark{3,6,7},  S. Seager\altaffilmark{3,6},  D. J. Osip\altaffilmark{8}, M. J. Holman\altaffilmark{2}, J. N. Winn\altaffilmark{6}, S. Hoyer\altaffilmark{9}, P. Rojo\altaffilmark{9}}

\altaffiltext{1}{This paper includes data gathered with the 6.5 meter Magellan Telescopes located at Las Campanas Observatory, Chile.}
\altaffiltext{2}{Harvard-Smithsonian Center for Astrophysics, 60 Garden St., Cambridge, MA, 02138}
\altaffiltext{3}{Department of Earth, Atmospheric, and Planetary Sciences, Massachusetts Institute of Technology, 77 Massachusetts Ave., Cambridge, MA, 02139}
\altaffiltext{4}{Institut de Ci\`encies de l'Espai (CSIC-IEEC), Campus UAB, Facultat de Ci\`encies, Torre C5, parell, 2a pl, E-08193 Bellaterra, Barcelona, Spain}
\altaffiltext{5}{Visiting Investigator; Carnegie Institution of Washington, Department of Terrestrial Magnetism, 5241 Broad Branch Road NW, Washington, DC 20015-1305}
\altaffiltext{6}{Department of Physics, Massachusetts Institute of Technology, 77 Massachusetts Ave., Cambridge, MA, 02139}
\altaffiltext{7}{Lowell Observatory, 1400 W. Mars Hill Rd., Flagstaff, AZ 86001}
\altaffiltext{8}{Las Campanas Observatory, Carnegie Observatories, Casilla 601, La Serena, Chile}
\altaffiltext{8}{Astronomy Department, Universidad de Chile, Casilla 36-D, Santiago de Chile, Chile}

\begin{abstract}

Although OGLE-TR-56b was the second transiting exoplanet discovered, only one light curve, observed in 2006, has been published besides the discovery data. We present twenty-one light curves of nineteen different transits observed between July 2003 and July 2009 with the Magellan Telescopes and Gemini South. The combined analysis of the new light curves confirms a slightly inflated planetary radius relative to model predictions, with $R_p = 1.378 \pm 0.090$ $R_J$. However, the values found for the transit duration, semimajor axis, and inclination values differ significantly from the previous result, likely due to systematic errors. The new semimajor axis and inclination, $a = 0.01942\pm 0.00015$ AU and $i = 73.72 \pm 0.18^{\circ}$, are smaller than previously reported, while the total duration, $T_{14} = 7931\pm38$ s, is 18 minutes longer. The transit midtimes have errors from 23 s to several minutes, and no evidence is seen for transit midtime or duration variations. Similarly, no change is seen in the orbital period, implying a nominal stellar tidal decay factor of $Q_* = 10^{7}$, with a three-sigma lower limit of $10^{5.7}$.

\end{abstract}


\section{Introduction}

OGLE-TR-56b was the second transiting exoplanet discovered, after HD209458b \citep{Charbonneau2000, Henry2000}, and the first found by a photometric wide-field survey \citep{Udalski2002b}. The planetary nature of this object was confirmed by \citet{Konacki2003a}, who derived a mass of $1.3 M_J$ from radial velocity measurements of the host star. An initial flurry of follow-up observations included radial velocity measurements by \citet{Torres2004} and \citet{Bouchy2005}; the determination of the host star's fundamental parameters and chemical composition by \citet{Santos2006}; and more recently the detection of OGLE-TR-56b's atmosphere by \citet{Sing2009}. However, due to the faintness of the target  ($V=16.56$, $I=15.3$) and the recent explosion in the number of other, brighter transiting systems, OGLE-TR-56b remains relatively unstudied, despite having one of the shortest known periods (1.2 days). The radius of the planet is currently constrained by only two light curves: the composite OGLE discovery light curve in $I$ \citep{Udalski2002b}, and a single light curve in 2006 observed alternately in Bessel $R$ and $V$ filters \citep{Pont2007}. The precision of the planetary parameters for this system has thus far been limited by the photometric quality of the available transit light curves \citep{Southworth2008} and not by the precision of the stellar parameters, as is currently the case for many other planets.  

OGLE-TR-56b was selected in 2006 as part of our campaign of high-quality photometric observations to search for transit timing variations \citep[see][]{Adams2010PhD,Adams2010a,Adams2010b,Adams2011}. Its short orbital period allows for many observational opportunities, and it remains one of the better candidates to search for changes in the orbital period due to orbital decay. Predictions vary widely on the time scale over which orbital decay would be observable \citep{Sasselov2003, Patzold2004, Carone2007, Levrard2009}, with estimates of the remaining lifetime of this system ranging from a few billion years \citep{Sasselov2003} to a much shorter 7 Myr \citep[assuming $Q_*=10^6$][]{Levrard2009} using very different assumptions about tidal equilibrium modes. Therefore, the estimates for the decrease in the orbital period due to orbital decay range from 0.1 to 10 ms yr$^{-1}$. This value is only a few times smaller than the current error on the orbital period \citep[86 ms from][]{Pont2007}. 

Here we present twenty-one new light curves of nineteen transits of OGLE-TR-56b, observed from 2003 to 2009, which we use to improve the planetary system parameters and also to search for potential timing anomalies in this system, including transit timing variations (TTVs) caused by interactions other bodies in the system, and transit duration variations (TDVs) caused by orbital precession, among other effects. In \S~\ref{section:ogle56obs} we describe the collection and analysis of the data. In \S~\ref{section:ogle56fitting} we describe the transit model fitting; we note the difference in light curve shape compared to the light curve published by \citet{Pont2007} in \S~\ref{section:ogle56shape}. In \S~\ref{section:ogle56timing} we present the revised transit ephemeris and discuss how the timing constrains the presence of other objects in the system.

\section{Observations and data analysis}
\label{section:ogle56obs}

Nineteen transits were observed between 2003 July and 2009 July, all but one with the Magellan telescopes. All transits are referred to by the UTC date of observations; when multiple light curves exist for a given UTC date, e.g. the two light curves on 2005 April 19 UT in the Johnson $B$ and $I$ band, they are denoted respectively as 20050419B and 20050419I. Seven full or partial transits were observed by the Transit Light Curve (TLC) collaboration \citep[e.g.,][]{Holman2006} during 2003-2006, using MagIC-SITe. Thirteen transits were observed between 2006-2009 on Magellan using three instruments: POETS (3 transits), IMACS (1 transit), and MagIC-e2v (9 transits). One additional transit was observed in 2009 using GMOS on Gemini South. Details on instrument setup and observational conditions are summarized in Table~\ref{table:ogle56obsparams}.

\begin{figure}
\includegraphics*[scale=0.65]{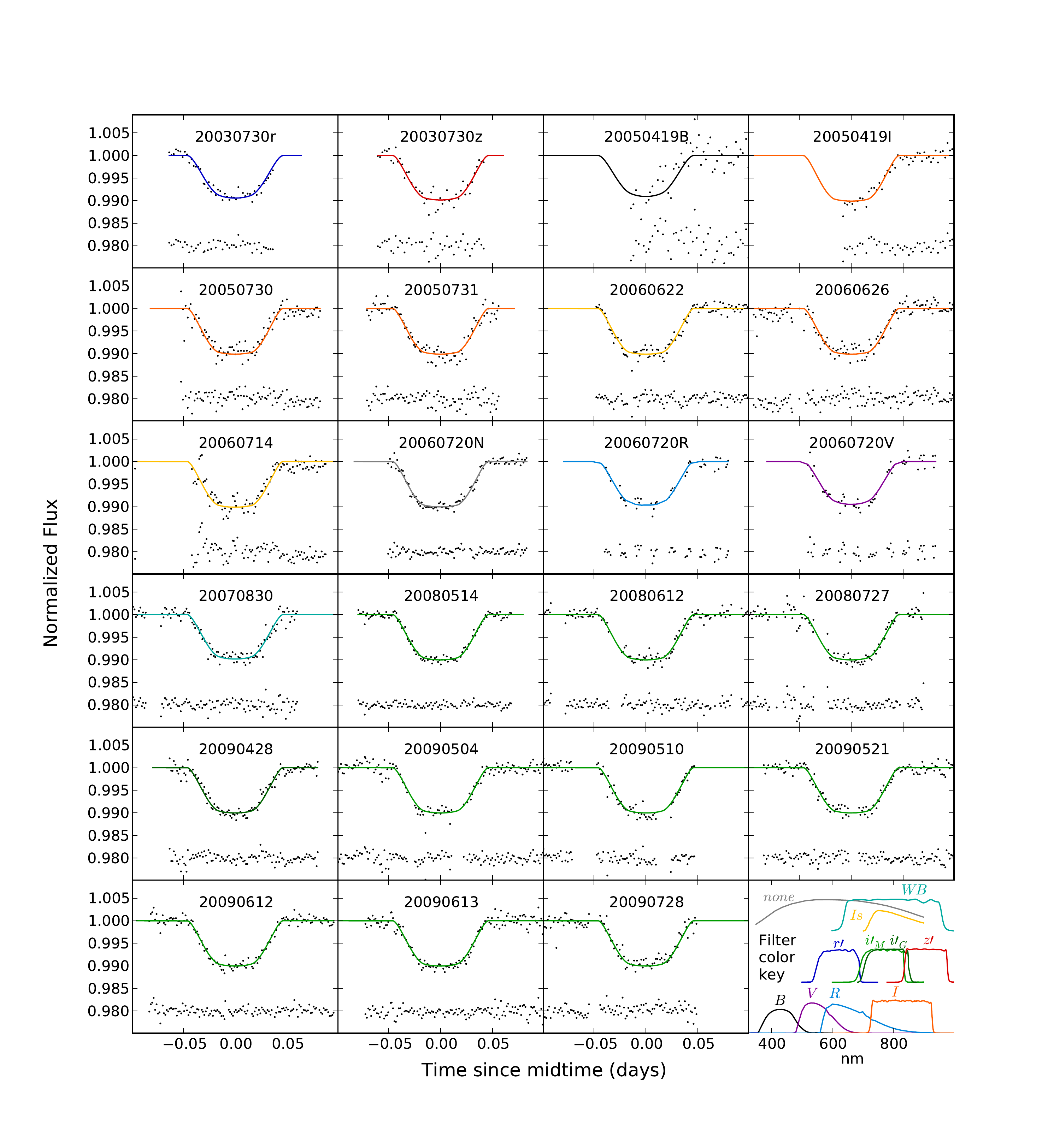}
\caption[Transits of OGLE-TR-56b]{Twenty-three light curves observed over nineteen different transit epochs of OGLE-TR-56b. All data are binned to 2 minutes to aid comparison. The solid lines show the best model for all curves, color-coded by the instrument and filter used: SITe/$B$ = black, FORS1/$V$=purple, SITe/$r'$ = dark blue,  FORS1/$R$=light blue, IMACS/WB6500-9300 = cyan, e2v/$i'$ = green, GMOS/$i'$ = dark green, POETS/$Is$ = yellow, SITe/$I$ = orange, SITe/$z'$ = red, and POETS/no filter = gray. The residuals from the model are plotted below each curve.}
\label{fig:ogle56transitsbinned}
\end{figure}

The Magellan instruments all had relatively small fields-of-view with high spatial sampling. The earliest transits used the Magellan Instant Camera (MagIC) with the original SITe CCD on the Clay telescope. MagIC-SiTe had a field of view of 140\arcsec\ $\times$ 140\arcsec\ and a plate scale of 0.069\arcsec\ per pixel;  the CCD gain was about 2 e-/ADU and the read noise 6 e-. Four filters were used: Harris $B$, Mould $I$, and Sloan $r'$ and $z'$ (the filter transmission curves are shown in the legend of Figure~\ref{fig:ogle56transitsbinned}).  In June and July 2006, POETS (Portable Occultation, Eclipse, and Transit System) was installed as a visiting PI instrument on the Clay telescope. POETS has a field of view of 23\arcsec\ $\times$ 23\arcsec\ and a plate scale of 0.\arcsec046 per pixel. The camera is a frame-transfer CCD which can be GPS triggered and is described in \citet{Souza2006}. POETS was operated full-frame (no binning) in conventional mode with a gain of 3.4 e-/ADU and read noise of 6 e-.  For the first two transits observed, a Schuler Astrodon Johnson-Cousins \emph{Is} filter was used. For the third transit, observed on 20 July 2006, coincidentally the same night as the transit by \citet{Pont2007}, no filter was used as a test to achieve greater throughput; the POETS CCD is sensitive to wavelengths between 400-900 nm \citep{Gulbis2008}. We refer to this as 20060720N to distinguish it from the \citet{Pont2007} observations.

One transit was observed in 2007 with the Inamori Magellan Areal Camera and Spectrograph (IMACS) on the Magellan Baade telescope \citep{Dressler2006}. We used the $f$/4 imaging mode with a subraster on one of the eight instrument CCDs (chip 2), reading out 1200x1200 pixels centered around x=1080 and y=3015, in order to decrease the readout time to roughly 35 seconds per frame in FAST mode. The area imaged was thus 133\arcsec\ $\times$ 133\arcsec, with a plate scale of 0.11\arcsec\ per pixel. This chip has a gain of 0.9 e-/ADU and read noise of 4.6 e-. A wideband filter, WB6300-9500, was used.

Between January 2008 and December 2009, the Magellan Instant Camera (MagIC) was mounted on the Magellan Baade telescope with the addition of a new e2v CCD. The MagIC-e2v detector, which shares a dewar with the older SITe CCD, is identical to the red CCD on HIPO (one of the first generation instruments to be flown on SOFIA), a fast read-out direct imaging camera that uses the LOIS control software \citep{Dunham2004, Taylor2004,Osip2008}. All of the MagIC-e2v transits were observed in the Sloan $i'$ filter. The MagIC-e2v camera has a field of view of 38\arcsec\ $\times$ 38\arcsec\ and a plate scale of 0.\arcsec037 per pixel unbinned.  The camera can be operated in two different modes: single exposure mode, with a readout time of about 5 seconds per exposure, which was used for the first three transits in 2008; and frame transfer mode, with a readout time of only 3 milliseconds between frames in an image cube, which was used for the following six transits. The gain and read noise of the transits observed in 2008 were 2.4 e-/ADU and 5.5 e- per pixel, respectively, and two amplifiers were used during readout; after July 2008, the CCD was reconfigured to have a gain of 0.54 e-/ADU and 5 e- read noise per pixel and a single readout amplifier.

The transits observed with the frame transfer camera POETS were observed at a very rapid cadence (2-6 s per exposure), in order to get good time sampling during transit ingress and egress. However, the noise per frame was such that we chose to bin every 10 frames (20 s) for 20060622, 8 frames (32 s) for 20060714, and 10 frames (60 s) for 20060720N. In subsequent transits, the exposure times were adjusted to maintain roughly $10^6$ integrated photons for both the target and multiple nearby comparison stars. Exposure times for unbinned ($1\times1$) data ranged from 17-60 sec, and for binned data from 10-60 sec. Details of the observing settings are noted in Table~\ref{table:ogle56obsparams}. Gaps in the data for transits 20060622, 20060714, and 20080727 were caused by separate instrument computer glitches, while the gap in transit 20090510 was due to observing a gamma ray burst for another project while OGLE-TR-56b passed through zenith.

The observation strategy for the transits observed by the Transit Light Curve collaboration between 2003-2006 is similar to that of OGLE-TR-111b. Details are provided in \citet{Winn2007}.

The transit on 20090428 was observed using GMOS (Gemini Multi-Object Spectrograph) at Gemini South Telescope. The observations were done in service mode using a Sloan $iÕ$ filter (\emph{i\_GO327}), $2\times2$ binning, and an integration time between 12 and 30 seconds. We used the Region of Interest (ROI) of the camera in order to reduce the readout time to only 24 seconds in fast readout mode and low gain. 309 frames of 1024x2304 pixels were obtained with a scale of 0.\arcsec146 per pixel. The chip has a gain of 5 e-/ADU and read noise of 7.8 e-.

Accurate timing is of the utmost importance for this project, so special care was taken to ensure that the correct times were recorded in the image headers. For the transits observed with POETS in 2006, each image frame was triggered by a GPS, so the UTC start times are accurate to the microsecond level. For the transits observed with GMOS, MagIC-SITe, IMACS and MagIC-e2v (in 2008), the UTC start times for each image were recorded from network time servers, which in most cases were verified by eye to be synchronized with the observatory's GPS clocks at the beginning of each night. For the MagIC-e2v observations in 2009, the times came from a small embedded control computer (a PC104), which received unlabeled GPS pulses every second. As with the network time servers, the PC104 was synchronized with the observatory's GPS before each transit observation. In all cases the UTC time signals written to the image headers agree within one second with the GPS time, as verified by examining the system control logs. 

\subsection{Data analysis}
\label{section:data}

All  data were calibrated using IRAF\footnote{IRAF is distributed by the National Optical Astronomy Observatories, which are operated by the Association of Universities for Research in Astronomy, Inc., under cooperative agreement with the National Science Foundation.}. The zero level for POETS and IMACS data frames were calculated from bias frames taken before or after the transits, while the MagIC data frames were corrected using overscan regions on each image. The images were flat-fielded using either dome or twilight flats in the appropriate filter, as available. The GMOS data were reduced using the Gemini pipeline.

\begin{figure}
\includegraphics*[scale=0.6]{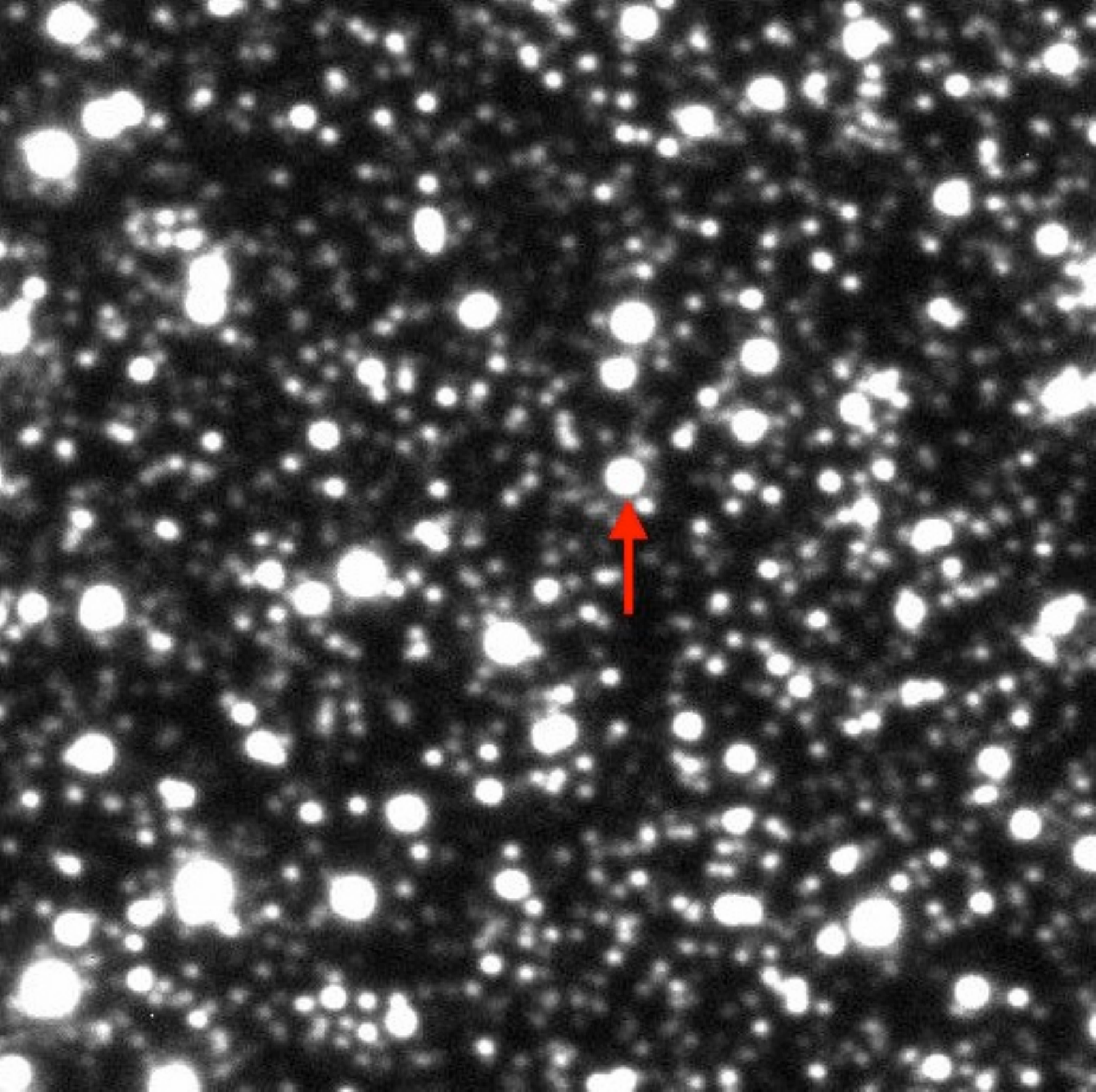}
\caption[OGLE-TR-56 on MagIC-e2v]{The field of view for OGLE-TR-56, as observed on 2008 May 14 with MagIC-e2v on the 6.5 m Baade telescope. Field of view is 38\arcsec\ x 38\arcsec, with north up and east to the left.}
\label{fig:ogle56fov}
\end{figure}

Although OGLE-TR-56 is in a very crowded field (Figure~\ref{fig:ogle56fov}), the generally excellent seeing and good spatial resolution of the detectors allowed for high-quality light curves using aperture photometry. Most of the photometry was done using the IRAF routine \emph{phot}, part of the \emph{apphot} package. A wide range of apertures and different comparison stars were examined, depending on the nightly conditions (e.g., seeing) and specific field of view, in the same way as described for other planets we have analyzed \citep{Adams2010a,Adams2010b,Adams2011}. 

For a few transits, an alternative aperture photometry method was used to achieve greater precision. In this method, boxes were drawn around the target and comparison stars, and the sky was determined using a single 30-pixel box drawn around an empty region of sky (not easy to find for targets in crowded fields, such as OGLE-TR-56b). This method, as implemented in \emph{Mathematica}, is both slower and not as robust as IRAF's \emph{phot}, particularly for data sets in which the stars shifted by more than a few pixels; but for two transits (20060622 and 20080514) it produced lower out-of-target scatter and cleaner light curves. The sky region for 20060622 was centered 2.1\arcsec west and 0.1\arcsec south of the transiting planet host star; for 20080514 the region was located 2.0\arcsec east and 0.3\arcsec south.

We additionally explored using image subtraction \citep{Alard1998, Alard2000} to create the light curve for one transit, 20090504. This test produced a light curve with slightly better scatter compared to the aperture light curve (1.0 vs 1.1 mmag in 2 min), but the depth of the image-subtracted light curve is much shallower ($k=0.092\pm0.002$) than the aperture light curve for either the individual fit for that light curve ($k=0.102\pm0.004$) or the joint fit ($k=0.1039\pm0.0004$). This problem is a known potential pitfall for image subtraction photometry \citep[see e.g. discussion in][]{Gillon2007,Adams2010a}. We note, however, that image subtraction does not always give erroneous depths; the TLC light curves were created with image subtraction \citep{Winn2007}, and the depths and other parameters, particularly of the full transits, agree well with light curves made with aperture photometry.

After obtaining light curves for each transit, the out-of-transit flux of each curve was examined for linear trends against several variables: airmass, seeing, telescope azimuth, $(x, y)$ pixel location, and time since beginning of transit. These parameters were chosen because they are either directly correlated with photometric trends (e.g. seeing, airmass), or are proxies for other effects that may be harder to measure (e.g. the telescope azimuth, which is a major component in the de-rotator rates, which were not recorded in the headers of all transit files). The data presented no significant systematics, except in three transits, for which we fit and remove linear trends for slight slopes in the flux over time (transits 20070830 and 20090504), and a slope with seeing (transit 20080727). Since these slopes were removed before the light curve parameter fitting step (see \S~\ref{section:ogle56fitting}), for simplicity, it is possible that covariances between those trends and the transit parameters might have a slight effect on the individual transit parameter. However, the joint fit values are unlikely to be affected since most of the transits fit had no slopes removed.

In the case of 20060720N, the light curve observed with no filter, we found that differential color extinction dominates the light curve systematics. Differential extinction, strongly correlated with airmass, is particularly troublesome, since it affects individual comparison stars differently. We therefore took a slightly different detrending approach for this dataset. The low airmass values (1.0 to 1.1) allow a linear approximation of the airmass term when detranding. In addition, we fit a separate linear slope to the ratio of the target to each comparison star, before attempting to combine them to produce the final transit light curves. Although we initially examined 13 different comparison stars over a wide range of photometric apertures, only two comparison stars could be used in the final light curve. We note that despite the relatively low scatter of the resulting light curve, there are dangers in this approach, notably in how to accurately detrend when there is little data on one side of the out-of-transit baseline to constrain the slope removed. However, given that the resulting light curve agrees well in depth and shape with the other, filtered transits, this method seems to have worked in in this case. Because correcting for differential extinction in bluer wavelengths introduces many systematic complications, we do not recommend observing planetary transits without filters as a general rule; if the goal is to achieve greater throughput, a redder wideband filter would be a better option.

The photometric parameters used to generate each light curve are shown in Table~\ref{table:ogle56photparams}. Each resulting light curve is plotted in Figure~\ref{fig:ogle56transitsbinned}, together with the residuals from the best model plot, with each transit binned to two minutes for comparison. The transit fluxes and times are available as an online table; an excerpt is shown in Table~\ref{table:ogle56data}. 

\subsubsection{Literature light curves}

The only published high-quality light curve for OGLE-TR-56b was observed on 2006 July 20 UT with the VLT in both $R$ and $V$, with alternating sequences in each filter of 7-8 exposures of 15-40 s each \citep{Pont2007}. Upon refitting that light curve (provided by F. Pont 2007, personal communication), we found that the shape of the transit does not agree well with the one obtained from the new light curves, most notably in an 18-minute discrepancy in the transit duration (see \S~\ref{section:ogle56shape}). 

In an attempt to resolve this problem, we downloaded the now publicly available on the VLT archive database raw image frames and re-analyzed the data for each filter (see Figures 1 and 3). Our new light curves, based on aperture photometry, are less precise than the original image subtraction light curve: 1.0 mmag in $R$-band and 1.2 mmag in $V$-band, compared to 0.7 mmag for the original combined V/R curve. However, the aperture light curves are consistent in transit depth and duration with the 19 other light curves presented in this paper. The revised curves, referred to as 20060720R and 20060720V henceforth, are used in all subsequent analyses, except as noted for comparison with the original photometry, which we call 20060720P. We also note that the times used in the \citet{Pont2007} light curve appear to be the start time, rather than the mid-exposure time of each frame, and thus are earlier than the corresponding points in our analysis by 8-20 s; however, this slight time offset is less than the midtime error on the transit, and cannot account for the duration difference.

We did not refit the OGLE survey light curve, which is a composite of 13 full and partial transits spanning several hundred nights \citep{Udalski2002b}. However, we do use the most up-to-date published midtime of that composite light curve \citep{Torres2004} in our timing analysis, after adding 66.184 s to correct for the UTC-TT offset (\S~\ref{section:ogle56timing}).

\section{Transit fitting results}
\label{section:ogle56fitting}

\subsection{Model}

Our transit model fits use the algorithm of \citet{Mandel2002} as implemented by \citet{Carter2009}, assuming white noise. We assumed that OGLE-TR-56b has zero obliquity, oblateness and orbital eccentricity. The stellar mass and radius values used to convert the model output into physical parameters were taken from \citet{Torres2008} and are listed in Table~\ref{table:ogle56mcmc}. We also fixed the orbital period to $P = 1.211909$ days \citep{Pont2007}, since that parameter is only used to convert $a/R_*$ into an orbital speed, and therefore has no major effect in the other model parameters. To model the stellar limb darkening, we assumed a quadratic law with initial values for the $u_1$ and $u_2$ parameters set to the ATLAS values for the appropriate filter \citep{Claret2000,Claret2004}\footnote{Initial values: $u_{1,none}=0.19$,  $u_{2,none}= 0.25$, $u_{1,B}=0.5458$,  $u_{2,B}= 0.2504$,  $u_{1,r'}= 0.4089$, $u_{2,r'}=0.2757$, $u_{1,z'}=0.2289$, $u_{2,z'}=0.3112$, $u_{1,I}=0.1958$, $u_{2,I}=0.3561$, $u_{1,i'}=u_{1,WB}=0.2146$, $u_{2,i'}=u_{2,WB}=0.3569$, $u_{1,V+R}=0.3157$, $u_{2,V+R}=0.35265$.} In all transits we found it necessary to fix the quadratic term $u_2$. In addition, we only fit for the linear term $u_1$ in the transits observed in the $i'$ and $I$-band filters, while leaving it fixed for the transits observed in the $B$, $r'$, $z'$, $WB$-bands and with no filter because the precision of these light curves is insufficient to constrain the value of $u_1$. The values for $u_1$ and $u_2$ are calculated using the\emph{jktld} program by \citet{Southworth2008}\footnote{http://www.astro.keele.ac.uk/~jkt/codes/jktld.html}, assuming the stellar parameter values $T=6119~K$, $\log{g}=4.2~\textrm{cm/s}^2$, $[M/H] = 0$, and $V_{micro} = 2$~km/s. We fixed the coefficients for the wideband transit 20070830 to be the same as the Sloan $i'$ filter, since the actual filter used, WB6300-9500, is centered at the same wavelength, though about twice as wide. The limb darkening parameters for the transit with no filter were derived by setting all other parameters fixed to a joint-fit value excluding this light curve, then fitting for $u_{1,none}$ and $u_{2,none}$ alone. 
\subsection{Light curve fits}

Each light curve was fit both independently and jointly with all other light curves using a Markov chain Monte Carlo (MCMC) method, using Gibbs sampling and Metropolis-Hastings stepping \citep{Tegmark2004, Holman2006}. This method is described in greater detail in e.g. \citet{Carter2011}. Three independent chains of a million links each were combined, discarding the first 50,000 links of each chain, to avoid any potential solution biases due to the initial values of the input parameters. For each fitted parameter -- the radius ratio, $k$, inclination, $i$, semimajor axis ratio, $a/R_*$ (global parameters for all transits), limb darkening coefficients $u_1$ and $u_2$ (global for each filter), and out-of-transit flux, $F_{OOT}$ and transit midtime, $T_C$ (specific to each transit) -- we find the posterior distribution, all of which are roughly Gaussian. The median value and 68.3\% confidence intervals of each parameter distribution are reported in Table~\ref{table:ogle56mcmc}. 

To account for excess correlated noise in the light curves, we calculated the time-averaged residuals. This is done by binning the residuals for each light curve into bins from 10 to 30 minutes, the typical time scales of correlated noise \citep{Pont2006}, and then calculating how much greater the actual noise is compared to the ideal noise assuming photon statistics. The excess noise varied from 1-2.8 times the predicted noise level, and has been included in the error bars on all transit midtime and individual parameters.

As a check that there were no additional slopes in the data that could affect the photometry, we ran a test MCMC fit to 12 light curves that included a slope with respect to airmass as an additional free parameter. None of the 12 full light curves fit in this way had a significant shift in transit parameters, so we did not included airmass slopes in the rest of our analysis.

Table~\ref{table:ogle56mcmc} summarizes the joint model fit values for the planet-star radius ratio, $k$, inclination, $i$, semimajor axis ratio, $a/R_*$, total transit duration, $T_{14}$, full transit duration, $T_{23}$, and ingress/egress duration, $T_{12}=T_{34}$. (The total transit duration is the time from first to last contact, while the full transit duration is the time when the disk of the planet is entirely overlapping the disc of the star. The ingress and egress duration are the times it takes for the transition, and are equal if the orbit is circular. See \citet{Winn2010} for a mathematical description of the transit nomenclature.) 

\begin{figure}
\includegraphics*[scale=0.8]{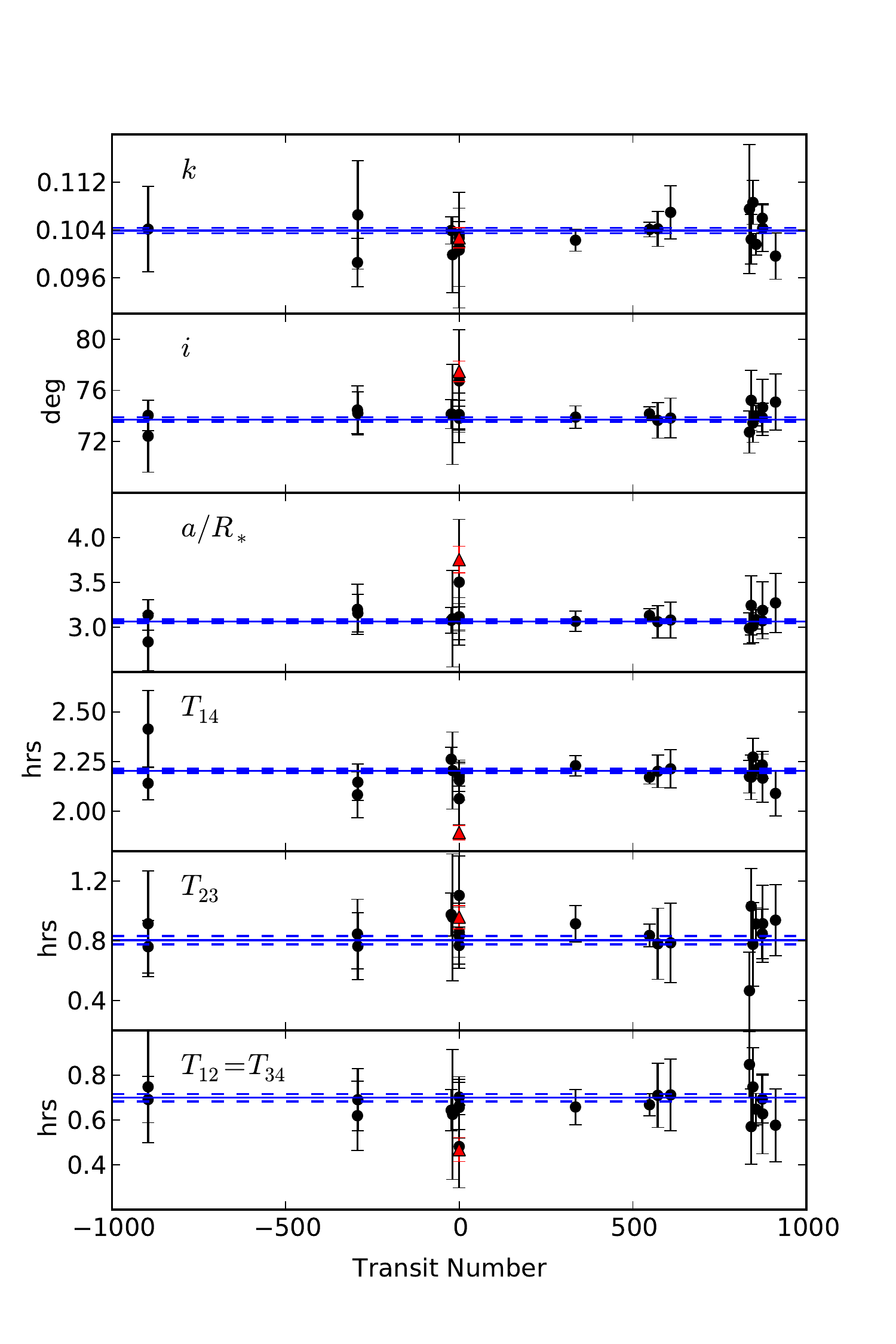}
\caption[Parameter variation of individual transits of OGLE-TR-56b]{Parameter variation of individual transits of OGLE-TR-56b. Data from the individual MCMC fits (Table~\ref{table:ogle56indfits}) is shown for the inclination ($i$), semimajor axis ($a/R_*$), planet-star radius ratio ($k$), total transit duration ($T_{14}$), full transit duration ($T_{23}$), and  ingress/egress duration, assuming a circular orbit ($T_{12}=T_{34}$). Note that all errors have been scaled upward based on the factor calculated from residual permutation. For comparison, the value derived from the joint MCMC fit is plotted as a solid blue lines with dashed $\pm1\sigma$ errors. We highlight our independent fit to the light curve published by \citet{Pont2007} in red to illustrate that it is markedly discrepant for all parameters except the radius ratio.}
\label{fig:ogle56paramvar}
\end{figure}

We also fit each light curve independently. Table~\ref{table:ogle56indfits} reports each individual parameter and error, along with the time-averaged residual factors by which those errors have been increased. To explore variations with time, the values are plotted in Figure~\ref{fig:ogle56paramvar}, with the joint-model fit shown as a horizontal line with 1-$\sigma$ errors.  A few partial or noisy light curves (20050419B, 20050419I, 20060714) have been omitted from the individual results; similarly, the radius ratio for 20030730z was fixed to the joint-fit value. In general, the independent fit resultss are consistent with the joint fit, with the notable exception of the original photometry for 20060720P, shown as a red triangle. No significant variations are seen in any of the studied parameters over the 6-year time span of the observations.

\subsection{Comparison to previous system parameters}
\label{section:ogle56shape}

With twenty-one new light curves, the combined fit provides a more precise determination of the system parameters, as summarized in Table~\ref{table:ogle56mcmc}. The errors derived from the photometry for the radius ratio, inclination, and semi-major axis are now significantly smaller than the error in the measured stellar radius; the component of the error in the planetary radius ratio due to stellar noise is fifteen times greater than the contribution from the photometry. Any improvement in the stellar parameters will significantly improve the precision with which the radius of OGLE-TR-56b is now known.

The radius ratio remains consistent with the value reported by \citet{Pont2007}, based on a single high-precision light curve. Moreover, of the nineteen light curves with independently fit radius ratio values in Table~\ref{table:ogle56indfits}, none lies more than 1.3 $\sigma$ from the joint-fit value. Most of the new light curves have radius ratio precisions ranging from 1-4\%. The lack of change in the transit depth over time indicates that either the star is less active than the few percent level, or that the pattern of star spots changes very slowly. No evidence was seen of star spot crossings.

\begin{figure}
\includegraphics*[scale=0.6]{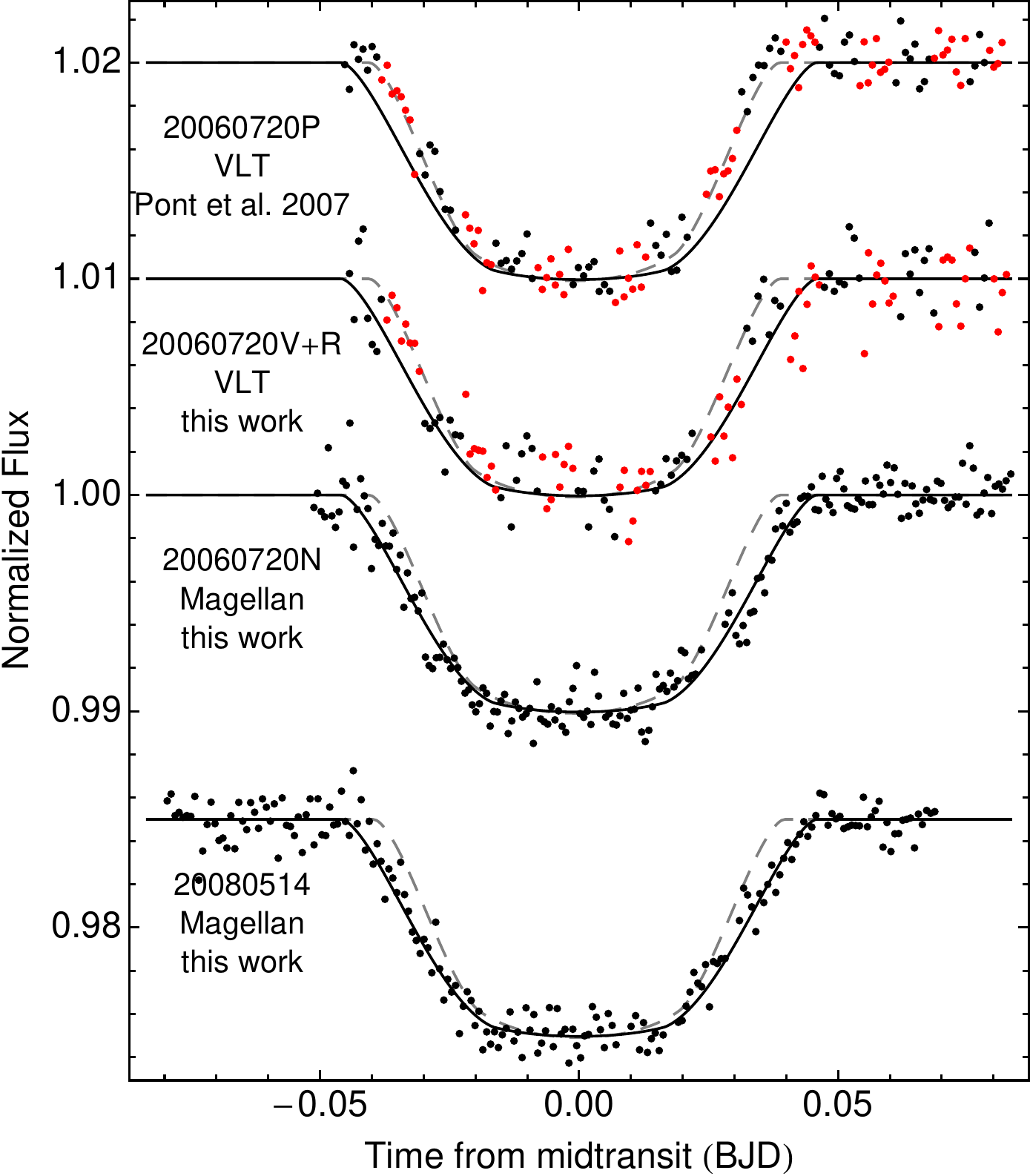}
\caption[Transit on 2006 July 20]{Three light curves of the transit on 2006 July 20, observed independently by two instruments, with a fourth curve on a different night for comparison. Data has not been binned. Top curve: original image-subtraction photometry for transit 20060720P observed with FORS1 on the VLT by \citet{Pont2007}, with $R$-band data in red and $V$-band data in black.  Second curve: the same VLT transit, redone with aperture photometry for this work. Third curve: independent observation of the same transit with no filter, 20060720N, using POETS on Magellan (discussed in \S~\ref{section:data}). The bottom curve is a different transit, 20080514, to illustrate the agreement of all but the top curve (20067020P) with the other transits presented in this work. Two models are shown in each plot: a gray dashed line for the best fit to the photometry by \citet{Pont2007} (refit for this work), and a black solid line showing the best joint-fit to all 23 new light curves. }
\label{fig:ogle56special}
\end{figure}

Although our radius ratio agrees with \citet{Pont2007}, the transit duration, $T_{14}$ is significantly longer, by 18 minutes, with corresponding differences also seen in the highly-correlated inclination and semi-major axis. The duration difference can be seen by eye, as is illustrated in Figure~\ref{fig:ogle56special} with three light curves of the same transit on 2006 July 20. The top curve in Figure~\ref{fig:ogle56special} corresponds to the original photometry reported in \citet{Pont2007} for the observations with the VLT. The second curve shows the same \citet{Pont2007} data frames, but with new aperture photometry in the same manner as the other transits reported in this paper. The third curve shows the same transit epoch observed independently on Magellan. The solid (black) and dashed (gray) lines in each panel show, respectively, the best fit to all new light curves, and the best fit to the \citet{Pont2007} light curve alone. A fourth curve for a different transit (20080514) is shown to illustrate the good agreement between the aperture photometry light curves and the rest of the data, though not the photometry of \citet{Pont2007}.

The duration of all transits, except for 20060720P, are best fit by the longer-lasting transit model (black solid line). We thus conclude that the photometry from \citet{Pont2007} most likely suffers from some unidentified noise source that shortens the apparent duration of the transit; this problem was likely exacerbated by the lack of data immediately before transit. This finding should be taken as a caution against using single-transit light curves to argue for transit duration variations due, for example, to planetary orbital precession.

The new value obtained for the semi-major axis ratio of the system, $a/R_{*} = 3.065 \pm 0.022$ versus the value of $3.75 \pm 0.15$ derived by \citet{Pont2007}, means that the planet is somewhat closer to the star than previously estimated. The implications for the planetary temperature are discussed in \S~\ref{section:ogle56temp}. The new value for the inclination, $i = 73.72 \pm 0.18$, also means that the planet is closer to a grazing orbit than previously thought ($b=0.859\pm0.003$).

We also note that the revised photometry should prompt a self-consistent reanalysis of the stellar parameters, which are now mildly discrepant. OGLE-TR-56 is faint (V = 16.56), and the radial velocity measurements have uncertainties from $20-100$ m $s^{-1}$ \citep{Torres2004, Bouchy2005}. Therefore the amplitude of the radial velocity signal, and subsequently the mass of the star, are not well constrained. \citet{Torres2008} derived a value of $M_* = 1.228 \pm 0.0078 M_{\odot}$, based on the stellar density derived from the photometry of \citet{Pont2007}. Combined with the orbital period, $P = 1.2119819$ days, this implies a semi-major axis of $3.76^{+0.35}_{-0.31}$, inconsistent with our new result ($a/R_{*} = 3.065 \pm 0.022$). The new photometry suggests a lower stellar density is required, with shifts of order $-1~\sigma$ in mass and $+3~\sigma$ in stellar radius needed to achieve consistent values for $a/R_*$.  A re-analysis of the the radial velocity data, perhaps including new measurements, are recommended to accurately determine the stellar parameters.

\subsection{Implications for the observed planetary occultation}
\label{section:ogle56temp}

\citet{Sing2009} obtained a $z'$-band occultation depth for OGLE-TR-56b of $0.0363 \pm 0.0091 \%$
and a brightness temperature of $T_{z'} = 2718^{+127}_{-107}$ K, using the values of the system parameters derived by \citet{Torres2008} and an effective temperature for the star of $6119 \pm 62$K. Since the system parameters have changed, we refit the occultation data in the same manner described in \S 3.1 of \citet{Sing2009}, replacing the orbital period, $a/R_*$, the planet-to-star radius ratio, and the orbital inclination by the new values in Table~\ref{table:ogle56mcmc}. The new fit gives an occultation depth of $0.0374\%$ and a central phase of $\phi = 0.4986 \pm 0.0007$. The central phase is within the errors of the previously derived value and still consistent with a circular orbit. The occultation is $0.0011\%$ deeper than the value derived before, but both values still agree within the measured depth uncertainty. The revised brightness temperature of the planet is $T_{z'} = 2708^{+102}_{-120}$ K, also consistent with the value derived before. Therefore, the atmospheric properties of the planet derived in \citet{Sing2009} remain valid, unless significantly different values for the stellar radius and temperature are found.

\section{Timing}
\label{section:ogle56timing}
 
With twenty different transit epochs of OGLE-TR-56b measured over eight years, the timing of the system can be quite well constrained. In Table~\ref{table:ogle56ominusc}, we present all available transit midtimes, including the OGLE survey time from \citet{Torres2004} as well as the new transits presented in this work and the re-analyzed data from \citet{Pont2007}. 

We calculate a new transit ephemeris using the transit times derived from the joint fit and also including the OGLE survey time (transit 20010615). Two transits observed in alternating filters (20030730 in $r'$ and $z'$ and 20060720 in $R$ and $V$) had divergent transit midtimes when split by filter, so we fit a single combined light curve for each transit individually (see also the combined fit parameter results in Table~\ref{table:ogle56indfits}). In both cases, the midtime from the combined light curve lies between the times derived from individual filter curves and is closer to the expected time of transit; this indicates the potential for timing errors if only a partial or poorly-sampled light curve is fit. (The individual filter curves for a third transit, 20050419, observed in two filters, $B$ and $I$, was left alone, because both filters cover only half a transit and yield large overlapping errors; although included in the weighted fit, this transit has minimal impact on the resulting ephemeris.) 

\begin{figure}
\includegraphics*[scale=0.6]{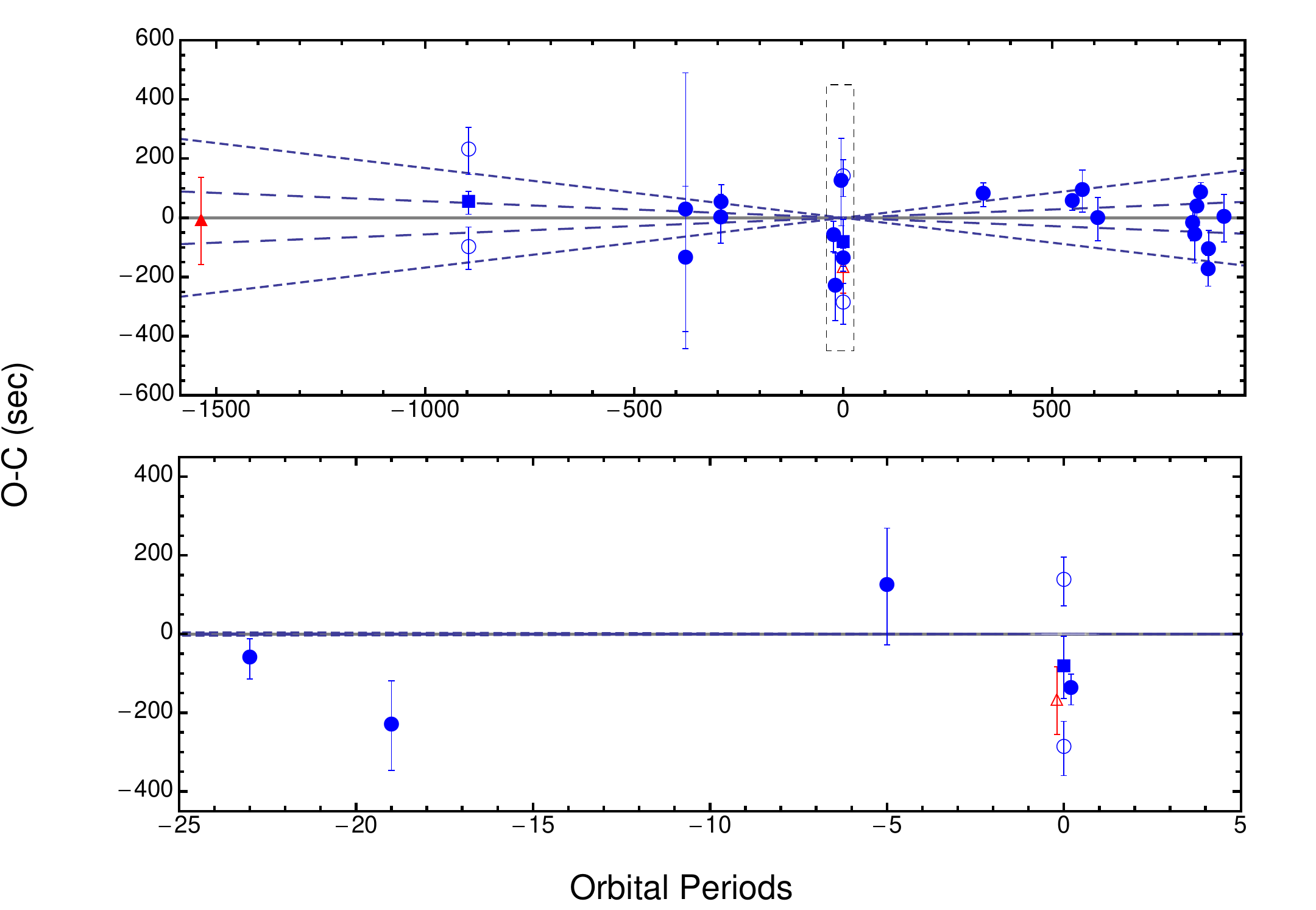}
\caption[Observed minus calculated midtimes for OGLE-TR-56b.]{Observed minus calculated midtimes for OGLE-TR-56b. Timing residuals using the new ephemeris (Equation~\ref{ogle56eqn1}). Solid symbols mark transits that were used to calculate the ephemeris. The solid red triangle at $-1536$ is the OGLE midtime, while the open red triangle at $0$ is our fit to the photometry from \citet{Pont2007}. The two solid blue squares at -896 and 0 represent independent fits to combined-filter light curves, which were used in place of the pair of open circles at those same values, representing each individual filter curve. See \S~\ref{section:ogle56timing} for more details. The bottom panel shows a zoomed-in  view of the region around 0. The solid line shows the expected time of transit, while the dashed lines represent the $1\sigma$ and $3\sigma$ errors on the calculated orbital period, indicating the slopes that result for a mis-determined period. Notice that the 1$\sigma$ and 3$\sigma$ error lines converge around the zero epoch and therefore appear practically indistinguishable in the lower plot.}
\label{fig:ogle56ominusc}
\end{figure}

We find the new transit ephemeris to be:
\begin{equation}
T_{C} = 2453936.60070(44) [BJD] + 1.21191096(65) N.
\label{ogle56eqn1}
\end{equation}
where $T_C$ is the predicted central time of a transit \citep[note that this is the same epoch as][]{Pont2007}, $N$ is the number of periods since the reference midtime, and the values in parentheses are errors on the last digits. The residuals from this ephemeris are shown in Figure~\ref{fig:ogle56ominusc}, with the lower panel zoomed in on the crowded region around the zero epoch. We note that the transit midtime is robustly measured for the zero epoch, with consistent timing results whether we use the original photometry (20060720P), the new photometry (20060720V+R) for the VLT light curve, or the independent time from Magellan (20060720N). The best linear fit has a reduced $\chi^2=2.6$, so we have scaled the errors on the parameters in Equation~\ref{ogle56eqn1} by $\sqrt{2.6}=1.6$. The relatively poor reduced $\chi^2$ suggests that the errors on some transits are still underestimated; removing the most discrepant 3-4 transits from the fit results in a reduced $\chi^2$ of just over 1. If these errors really are correct, then it is possible that the assumption of a constant ephemeris is incorrect.  The most deviant points are 20060720N at transit number 0  ($-3.6~\sigma$), and 20090612 at transit number 873 ($-3.4~\sigma$), but there is no reason to conclude at this time that intrinsic timing variations are responsible.

\subsection{Limits placed on timing variations}
\label{section:ogle56masslimits}

The error on the revised period estimate is 56 ms, with no evidence of decrease seen over the eight year time span from 2001-2009. If we fit for a linear decrease in the orbital period over time, the rate in change of the period is $\dot{P}=-2.9\pm17$ ms yr$^{-1}$, consistent with a constant period. Based on this result, we can derive a conservative lower limit estimate for the value of the stellar tidal decay factor, $Q_*$. Assuming a three-sigma upper limit on the period change (5.4 ms yr$^{-1}$) and Equation 5 from \citet{Levrard2009}, we find that the tidal decay factor for OGLE-TR-56 is no less than $Q_* = 10^{5.7}$. Furthermore, the nominal value obtained for the period change implies a $Q_* = 10^{7}$. 

No evidence is seen of transit duration variations. The best fit to the individual transit durations listed in Table~\ref{table:ogle56indfits}, using the combined light curves 20030730r+z and 20060720R+V in place of the curves separated by filter, and omitting the much-shorter duration light curve 20060720P, finds that the total transit duration has changed by $\dot{T_{14}}=-0.3\pm32$ s yr$^{-1}$, consistent with no change. Since the planet has a large impact parameter and nearly-grazing transits, it is possible that in the future orbital precession might lead to observable changes in the planetary duration. However, no such changes have been seen to date.

We performed numerical integrations to place limits on the perturber mass for a subset of the data, using the same method for other planets described in \citet{Adams2010a,Adams2010b,Adams2011}. Using 11 transits from 2006-2009, companion mass limits were placed down to $12 M_J$ in the exterior 2:1 mean motion resonance. However, we were unable to successfully run the analysis on the full set of transit times, most likely owing either to intrinsic instability in the system (meaning very few potential companions would be stable) or an error in our assumptions (that the timing residuals are intrinsically flat). Given the high reduced $\chi^2$ value of the constant-period fit in Equation~\ref{ogle56eqn1}, it is most likely that either a few transit errors remain underestimated, or that there are timing variations that will require more data to conclusively identify and characterize.

\section{Conclusions}
\label{section:ogle56conclusions}

In this work we have presented 21 new light curves of 19 transits of OGLE-TR-56b, vastly increasing the supply of high quality data on the planet. Our fitted radius value of $1.378\pm0.090~R_J$  is almost identical to the previously published value, and we note that the error is almost entirely supplied by error on the stellar radius; the component of uncertainty from the photometry alone is fifteen times smaller.

The values presented for the transit duration, inclination, and semi-major axis in this work are significantly different from those reported earlier, the most likely explanation being an error in the previous photometry. The new value for $a/R_*=3.065 \pm 0.022$ places the planet slightly closer to its star than previously thought, while the inclination, $i= 73.72\pm0.18$ degrees, means that planet is closer to a grazing orbit ($b=0.859\pm0.003$). Furthermore, the photometrically-derived values for the semi-major axis ratio are somewhat inconsistent with the current values for the stellar radius and mass; the stellar radius would need to be larger and the stellar mass smaller in order for the values to agree. New measurements and analysis of the stellar parameters are needed to resolve this inconsistency. In particular, if the stellar radius really is larger, the planetary radius value would also increase, implying that OGLE-TR-56b has an even more inflated radius than currently thought, with implications for its composition and energy budget.

The new orbital period is $P=1.21191096 \pm 0.00000065$ days. Although there are a few points with 3-sigma residuals, there is no reason to conclude at this time that these are transit timing variations, the most likely explanation being underestimated errors. Excluding the likely erroneous low value for the previously published duration, the transit duration has remained constant during a nine year span (2001-2009), with the best fit change $\dot{T_{14}}=-0.3\pm32$ s yr$^{-1}$. The orbital period has likewise been constant, with $\dot{P}=-2.9\pm17$ ms yr$^{-1}$, consistent with no change. Taking the three-sigma upper limit on the period change (5.4 ms yr$^{-1}$) means we can place a lower limit on the stellar tidal decay factor of $Q_* = 10^{5.7}$; the nominal value implies $Q_* =10^7$.

\acknowledgements

E.R.A. received support from NASA Origins grant NNX07AN63G. M.L.M. acknowledges support for parts of this work from NASA through Hubble Fellowship grant HF-01210.01-A/HF-51233.01 awarded by the STScI, which is operated by the AURA, Inc. for NASA, under contract NAS5-26555. This paper makes use of observations made with the European Southern Observatory telescopes and obtained from the ESO/ST-ECF Science Archive Facility.We thank Paul Schechter for observing a transit of OGLE-TR-56b as part of MIT's Magellan queue; Brian Taylor and Paul Schechter for their tireless instrument support; and Georgi Mandushev for assistance with image subtraction.




\newpage

\begin{deluxetable}{llllllllllll}
\tablewidth{0pt}
\tabletypesize{\scriptsize}
\tablecaption{Observational details for twenty-one new light curves of OGLE-TR-56b}
\tablehead{
Transit		& Instrument	& Frames\tablenotemark{a}	& Exposure	& Filter	& Binning	& Read 	& Airmass 	& Time observed		& Atm.	& Seeing	\\
(UT)			&			& 						& (sec)		&		& 		& (sec)		&	 		& (before ingress)	& stability & (arcsec)}
\startdata
20030730r	& MagIC-SITe	& 47 (0)					& --		&$r'$			& 1x1	& 23				& 1.1--1.7	 & 2.4 hrs (+27 min)	& --		&-- \\
20030730z	& MagIC-SITe	& 48 (0					& --		&$z'$		& 1x1	& 23				& 1.1--1.7	 & 2.4 hrs (+22 min)	& --		&-- \\
20050419B	& MagIC-SITe	& 59 (0)					& --		& B			& 1x1	& 23				& 1.0--1.6	 & 3.2 hrs ($-54$ min)& --		&-- \\
20050419I	& MagIC-SITe	& 59 (0					& --		& I			& 1x1	& 23				& 1.0--1.6	 & 3.3 hrs ($-53$ min)& --		&-- \\
20050730		& MagIC-SITe	& 140 (0)					& --		& I			& 1x1	& 23				& 1.0--1.4	 & 3.2 hrs (+10 min)	& --		&-- \\
20050731		& MagIC-SITe	& 113 (0)					& --		& I			& 1x1	& 23				& 1.1--1.9	 & 3.0 hrs (+37 min)	& --		&-- \\
20060622		& POETS		&6642 (121)\tablenotemark{b}	& 2		& Is			& 1x1	&\tablenotemark{c}	& 1.0--1.4	 & 3.9 hrs (+13 min)	& Stable		& 0.25-0.5 \\
20060626		& MagIC-SITe	&245 (0)					& --		& I			& 1x1	& 23				& 1.0--2.3	 & 4.6 hrs (+71 min)	& -		&-- \\
20060714		& POETS		& 2200 (588)\tablenotemark{d}	& 4		& Is			& 1x1	&\tablenotemark{c}	& 1.0--1.4	 & 4.6 hrs (+88 min)	& Variable	& 0.25-0.5  \\
20060720N	& POETS		&1867 (1379)\tablenotemark{e}& 6		& None		& 1x1	&\tablenotemark{c}	& 1.0--1.1	 & 5.5 hrs (+17 min)	& Stable		& 0.4  \\
20070830		& IMACS		&352 (1)\tablenotemark{f}		& 15, 30	& WB\tablenotemark{g}	& 1x1	& 25		& 1.0--1.5 	 & 4.4 hrs (+112 min)& Stable		& 0.6 \\
20080514		& MagIC-e2v	&187 (0)					& 60		&$i'$			& 1x1	& 5				& 1.0--1.5	 & 3.6 hrs (+56 min)	& Superb		& 0.3-0.4 \\
20080612		& MagIC-e2v	& 604 (64)\tablenotemark{h}	& 17, 20	&$i'$			& 1x1	& 5				& 1.0--1.2	 & 7.9 hrs (+264 min)& Stable		& 0.3-0.5 \\
20080727		& MagIC-e2v	& 672 (0)\tablenotemark{i}	& 10,12	&$i'$			& 2x2	& 5				& 1.0--1.3	 & 4.3 hrs (+100 min)& Stable		& 0.7 \\
20090428		& GMOS		& 309 (0)					& 12-30	&$i'$			& 2x2	& 24				& 1.0--1.5	 & 3.4 hrs (+32 min)	& Stable		& 0.4-0.6 \\
20090504		& MagIC-e2v	& 623 (2)\tablenotemark{i}	& 23, 25	&$i'$			& 2x2	&\tablenotemark{c}	& 1.0--1.2	 & 5.5 hrs (+90 min)	& Variable	& 0.5-1.0 \\
20090510		& MagIC-e2v	& 460 (2)\tablenotemark{f}	& 20, 30	&$i'$			& 2x2	&\tablenotemark{c}	& 1.0--1.2	 & 3.7 hrs (+94 min)	& Stable		& 0.7 \\
20090521		& MagIC-e2v	& 626 (0)					& 22-25	&$i'$			& 2x2	&\tablenotemark{c}	& 1.0--1.2	 & 4.6 hrs (+65 min)	& Variable	& 0.4-0.7 \\
20090612		& MagIC-e2v	& 772 (3)\tablenotemark{f}	& 17, 20	&$i'$			& 2x2	&\tablenotemark{c}	& 1.0--2.3  & 4.3 hrs (+63 min)	& Stable		& 0.4-0.7 \\
20090613	&MagIC-e2v	& 715 (0)					& 20		&$i'$			& 2x2	&\tablenotemark{c}	& 1.0--2.0	 & 4.3 hrs (+48 min)	& Variable	& 0.6-0.8 \\
20090728	&MagIC-e2v	& 535 (5)\tablenotemark{f}	& 30-60	&$i'$			& 2x2	&\tablenotemark{c}	& 1.0--1.2	 & 5.3 hrs (+187 min)&  Variable	& 0.4-0.9 \\
\enddata
\tablenotetext{a} {Number of frames used (additional frames that were discarded).}
\tablenotetext{b} {Discarded during meridian crossing (images elongated).}
\tablenotetext{c} {Readout is a few miliseconds in frame transfer mode.}
\tablenotetext{d} {Discarded baseline after transit at a different pointing, which did not return to the same ratio level as before.}
\tablenotetext{e} {Discarded 31 frames during meridian crossing, 1347 frames at airmass $>$ 1.1, and 1 frame with an aberrant ratio.}
\tablenotetext{f} {Discarded a few anomalous frames, e.g. due to a seeing spike on one or more frames}
\tablenotetext{g} {WB6300-9500, a wide band filter from 630-950 nm.}
\tablenotetext{h} {Discarded initial frames with low counts on target (50 points) and a minor seeing spike (14 points).}
\tablenotetext{i} {Discarded frames with large target diameters (poor seeing): $>$11.5 pixel (20080727) and $>$11.0 pixel (20090504)}
\label{table:ogle56obsparams}
\end{deluxetable}

\begin{deluxetable}{lclllllllll}
\tablewidth{0pt}
\tabletypesize{\scriptsize}
\tablecaption{Aperture photometry parameters for each transit}
\tablehead{
Transit		& Comp.	& Aperture\tablenotemark{a}	& Sky radius, width	& Slope removed 	& Scatter\tablenotemark{b}	\\
(UT)			& Stars	&  (pixels)					& (pixels)			&				& (mmag)				} 
\startdata
20060622		& 5		& 10\tablenotemark{c}	& --					& --							& 1.0		\\
20060714		& 2		& 8.2					& 30, 10				& --							& 1.7		\\
20060720R	& 6		& 6.0					& 50, 5				& --							& 1.0		\\
20060720V	& 5		& 7.0					& 40, 5				& --							& 1.2		\\
20060720N	& 2		& 14					& 70, 5			&$-3.6\%$, $-9.6\%$ \tablenotemark{d}	& 0.7		\\
20070830		& 3		& 7.6					& 20, 20				& $-0.96\%$ \tablenotemark{e}		& 1.1		\\
20080514		& 5 		& 14.5\tablenotemark{c}	& --					& --							& 0.6		\\
20080612		& 5		& 12.2				& 50, 10				& --							& 0.9		\\
20080727		& 3		& 9.0					& 25, 30				& $+0.16\%$ \tablenotemark{f}		& 1.3		\\
20090428		& 4		& 4.0					& 30,	 5				& --							& 1.0		\\
20090504		& 2		& 8.2					& 20,	 30				& $+1.48\%$ \tablenotemark{e}	& 1.1		\\
20090510		& 3		& 7.8					& 40, 10				& --							& 1.0		\\
20090521		& 5		& 7.0					& 30, 10				& --							& 0.9		\\
20090612		& 5		& 8.2					& 30, 10				& --							& 0.8		\\
20090613		& 3		& 10.2				& 25, 10				& --							& 1.1		\\
20090728		& 2		& 8.2					& 15,	 20				& --							& 1.2		\\
\enddata
\tablenotetext{a} {Except as noted, the aperture is a circular radius around the star.}
\tablenotetext{b} {Scatter on out-of-transit flux, binned to 2 minutes.}
\tablenotetext{c} {Box half-width for alternate photometry method described in \S~\ref{section:data}}
\tablenotetext{d} {Trend removed for each star, individually, against airmass, in units of Z day$^{-1}$.}
\tablenotetext{e} {Trend removed against time, in units of flux day$^{-1}$.}
\tablenotetext{f} {Trend removed against seeing in pixels, in units of flux pixels$^{-1}$.}
\label{table:ogle56photparams}
\end{deluxetable}

\begin{deluxetable}{l l l l }
\tablewidth{0pt}
\tabletypesize{\scriptsize}
\tablecaption{Flux values for new transits of OGLE-TR-56b\tablenotemark{a}}
\tablehead{ Mid-exposure ($JD_{UTC}$)	& Mid-exposure ($BJD_{TDB}$)	& \textrm{Flux}  		& Error}
\startdata
2453908.670845	&	2453908.677416	&	1.002203		&	0.001641 \\
2453908.670868	&	2453908.677439	&	1.007551		&	0.001641\\
2453908.670891	&	2453908.677462	&	0.9947374	&	0.001641\\
2453908.670914	&	2453908.677485	&	1.003771		&	0.001641\\
2453908.670937	&	2453908.677508	&	0.9926688	&	0.001641\\
\nodata
\enddata     
\tablenotetext{a} {Full table available online.}
\label{table:ogle56data}
\end{deluxetable}

\begin{deluxetable}{l l }
\tablewidth{0pt}
\tabletypesize{\scriptsize}
\tablecaption{New system parameters for OGLE-TR-56}
\tablehead{Parameter & \textrm{Value}\tablenotemark{a}	}
\startdata
Radius ratio, $k$							& $0.1039 \pm 0.0004$ 				\\
Semimajor axis ratio, $a/R_*$					& $3.065 \pm 0.022$ 				\\
Inclination,  $i$ (deg)						& $73.72\pm0.18$ 			\\
Eccentricity,  $e$ 							& 0 (fixed)  			\\
Linear limb-darkening coefficient, $u_{1,i'}$		& $0.169 \pm 0.033$ 				\\
Quadratic limb-darkening coefficient, $u_{1,I}$ 	& $0.175 \pm 0.041$ 				\\
Impact parameter, $b$						& $0.859\pm0.003$ \\
Total transit duration, $T_{14}$ (sec)	 		& $7931\pm38$ \\
Duration of full transit, $T_{23}$ (sec)	 		& $2896^{+106}_{-100}$				 \\
Ingress/egress duration, $T_{12}= T_{34}$ (sec)	& $2518^{+59}_{-62}$					 \\
Orbital period, $P$ (days)						&$1.21191096 \pm 0.00000065$	\\
Reference epoch, $T_0$ (BJD$_{TDB}$)			& $2453936.60070 \pm0.00044$		\\
Semimajor axis, $a$ (AU)				 	 	& $0.01942\pm0.00015$\tablenotemark{b} \\
Planetary radius, $R_p$ ($R_J$)			 	& $1.378 \pm 0.090$\tablenotemark{b} \\
Planetary mass, $M_p$ ($M_{J}$)				& $1.39\pm0.18$\tablenotemark{b} \\
Stellar radius, $R_*$ ($R_{\odot}$)				& $1.363 \pm 0.089$\tablenotemark{b} \\
Stellar mass, $M_*$ ($M_{\odot}$)				& $1.228\pm0.078$\tablenotemark{b} \\
 \enddata     
\tablenotetext{a} {Median value of parameter distribution from joint fit to 23 light curves, with errors reported from the $68.3\%$ credible interval.}
\tablenotetext{b} {Stellar mass and radius values from \citet{Torres2008}.}
\label{table:ogle56mcmc}
\end{deluxetable}

\begin{deluxetable}{l c c c c c c c}
\tablewidth{0pt}
\tabletypesize{\scriptsize}
\tablecaption{Parameters for independent fits to each transit of OGLE-TR-56b}
\tablehead{
Transit 		&TAR\tablenotemark{a}		&$k$   			&$i$ (deg)		&$a/R_*$			&$T_{14}$		&$T_{23}$	&$T_{12}=T_{34}$  }
\startdata
20030730r	& 1.7	& $0.10414\pm0.0072$	& $72.43\pm2.8$	& $2.836\pm0.32$	& $8688.0\pm696.0$	& $3292.1\pm1271.0$	& $2693.0\pm895.0$ \\
20030730z	& 1.0	& 0.1039 (fixed)	& $74.05\pm1.2$	& $3.135\pm0.17$	& $7707.3\pm298.0$	& $2737.2\pm637.0$	& $2489.3\pm372.0$ \\
20050730r+z\tablenotemark{b} & 1.6	& $0.09856\pm0.004$	& $74.47\pm1.9$	& $3.199\pm0.28$	& $7500.7\pm417.0$	& $3044.2\pm836.0$	& $2229.3\pm558.0$ \\
20050731		& 1.3	& $0.10653\pm0.0091$	& $74.21\pm1.7$	& $3.156\pm0.21$	& $7728.1\pm330.0$	& $2749.0\pm804.0$	& $2485.9\pm499.0$ \\
20060622		& 1.9	& $0.10392\pm0.0023$	& $74.15\pm1.1$	& $3.077\pm0.14$	& $8143.4\pm216.0$	& $3510.9\pm516.0$	& $2317.3\pm334.0$ \\
20060626		& 2.1	& $0.09986\pm0.0064$	& $74.12\pm3.9$	& $3.095\pm0.54$	& $7938.0\pm696.0$	& $3446.9\pm1528.0$	& $2250.2\pm1046.0$ \\
20060720P\tablenotemark{c}& 1.0	& $0.10269\pm0.0017$	& $77.48\pm0.8$	& $3.754\pm0.15$	& $6817.9\pm129.0$	& $3452.3\pm281.0$	& $1684.0\pm192.0$ \\
20060720R	& 1.1	& $0.10064\pm0.0097$	& $73.85\pm1.9$	& $3.094\pm0.24$	& $7764.9\pm361.0$	& $3001.0\pm783.0$	& $2362.6\pm492.0$ \\
20060720V	& 1.1	& $0.1011\pm0.0066$	& $76.74\pm4.0$	& $3.503\pm0.7$	& $7428.3\pm475.0$	& $3973.5\pm948.0$	& $1733.8\pm664.0$ \\
20060720V+R\tablenotemark{d}& 1.3	&0.1039 (fixed)	& $73.82\pm0.9$	& $3.092\pm0.14$	& $7822.6\pm265.0$	& $2766.6\pm449.0$	& $2528.1\pm283.0$ \\
20060720N	& 1.6	& $0.1033\pm0.0021$	& $74.12\pm1.1$	& $3.116\pm0.15$	& $7858.5\pm204.0$	& $3088.4\pm604.0$	& $2385.5\pm380.0$ \\
20070830		& 1.0	& $0.10228\pm0.0018$	& $73.91\pm0.9$	& $3.065\pm0.11$	& $8025.8\pm182.0$	& $3290.0\pm439.0$	& $2369.0\pm283.0$ \\
20080514		& 1.1	& $0.10408\pm0.0012$	& $74.17\pm0.5$	& $3.131\pm0.07$	& $7817.2\pm121.0$	& $3008.5\pm276.0$	& $2405.0\pm180.0$ \\
20080612		& 1.8	& $0.10419\pm0.0029$	& $73.66\pm1.4$	& $3.061\pm0.18$	& $7924.2\pm293.0$	& $2807.6\pm857.0$	& $2556.6\pm517.0$ \\
20080727		& 1.7	& $0.10695\pm0.0045$	& $73.84\pm1.6$	& $3.08\pm0.2$	& $7968.5\pm346.0$	& $2831.0\pm961.0$	& $2563.7\pm576.0$ \\
20090428		& 1.7 & $0.10751\pm0.0108$ 	& $72.74\pm1.6$ 	& $2.987\pm0.17$ 	& $7829.0\pm296.0$ 	& $1676.7\pm935.0$ 	& $3051.5\pm528.0$ \\
20090504		& 2.6	& $0.10244\pm0.0042$	& $75.23\pm2.3$	& $3.243\pm0.33$	& $7817.5\pm404.0$	& $3712.0\pm905.0$	& $2053.5\pm605.0$ \\
20090510		& 2.0	& $0.10861\pm0.0037$	& $73.46\pm1.5$	& $3.013\pm0.19$	& $8182.1\pm340.0$	& $2795.2\pm1012.0$	& $2691.6\pm630.0$ \\
20090521		& 1.3	& $0.10159\pm0.0018$	& $74.01\pm0.8$	& $3.085\pm0.1$	& $7954.7\pm171.0$	& $3286.5\pm391.0$	& $2335.0\pm253.0$ \\
20090612		& 1.9	& $0.10597\pm0.0025$	& $73.88\pm1.1$	& $3.07\pm0.15$	& $8039.0\pm245.0$	& $3048.9\pm598.0$	& $2496.1\pm387.0$ \\
20090613		& 1.9	& $0.10429\pm0.0039$	& $74.67\pm2.2$	& $3.188\pm0.32$	& $7801.5\pm433.0$	& $3291.9\pm926.0$	& $2258.6\pm641.0$ \\
20090728		& 2.1	& $0.09963\pm0.0039$	& $75.09\pm2.2$	& $3.271\pm0.33$	& $7524.5\pm408.0$	& $3377.5\pm857.0$	& $2076.4\pm585.0$ \\

\enddata
\tablenotetext{a} {Multiplicative factor indicating excess noise, as derived from the time-averaged residual method.}
\tablenotetext{b} {Combined $r'$ and $z'$ photometry.}
\tablenotetext{c} {New fit to original photometry of \citet{Pont2007}}
\tablenotetext{d} {Combined $V$ and $R$ photometry.}
\label{table:ogle56indfits}
\end{deluxetable}

\begin{deluxetable}{l r r r r r}
\tablewidth{0pt}
\tabletypesize{\scriptsize}
\tablecaption{OGLE-TR-56b Transit Midtimes and Residuals}
\tablehead{Transit	 		& Number		& Midtime ($BJD_{TDB}$)			& O-C (s)&  		$\sigma$	& Source}
\startdata
20010615\tablenotemark{a}	& -1536 		& $ 2452075.10534 \pm 0.00170$		&$ -11 \pm 147$ 	& -0.1	& \citet{Torres2004} \\
20030730r				& -896 		& $ 2452850.73111 \pm 0.00092$		&$ 226\pm 80$		&  2.9 	&this work\\
20030730z				& -896 		& $ 2452850.72730 \pm 0.00083$		&$-103 \pm 72$	& -1.4 	&this work\\
20030730r+z\tablenotemark{b}& -896 		& $ 2452850.72908 \pm 0.00045$		&$ 51 \pm 39$		&  1.4 	&this work\\
20050419B				& -377 		& $ 2453479.71055 \pm 0.00540$		& $ 24 \pm 466$ 	&  0.1 	&this work\\
20050419I				& -377 		& $ 2453479.70867 \pm 0.00290$		& $-139 \pm 246$	& -0.6 	&this work\\
20050730					& -293 		& $ 2453581.51076 \pm 0.00096$		& $-3 \pm 83$		& -0.0 	&this work\\
20050731					& -292 		& $ 2453582.72328 \pm 0.00073$		& $ 49 \pm 63$		&  0.8 	&this work\\
20060622					& -23 		& $ 2453908.72603 \pm 0.00060$		& $-63 \pm 51$		& -1.2 	&this work\\
20060626					& -19 		& $ 2453913.57171 \pm 0.00130$		& $-233 \pm 114$	& -2.0 	&this work\\
20060714					& -5 			& $ 2453930.54256 \pm 0.00170$		& $121\pm 148$	&  0.8 	&this work\\
20060720P\tablenotemark{c}	& 0 			& $ 2453936.59875 \pm 0.00100$		&$-169 \pm 86	$ 	& -1.9 	& \citet{Pont2007} \\
20060720V				& 0 			& $ 2453936.59735 \pm 0.00080$		&$-291 \pm 69	$ 	& -4.2 	& this work\\
20060720R				& 0 			& $ 2453936.60226 \pm 0.00072$		&$ 134 \pm 62	$ 	&  2.2 	& this work\\
20060720V+R\tablenotemark{d}& 0 		& $ 2453936.59973 \pm 0.00092$		&$-85 \pm 79	$ 	& -1.1 	& this work\\
20060720N				& 0 			& $ 2453936.59908 \pm 0.00045$		&$-141 \pm 39	$ 	& -3.6 	& this work\\
20070830					& 335		& $ 2454342.59179 \pm 0.00046$		& $78 \pm 40$		& 2.0 	&this work\\
20080514					& 548 		& $ 2454600.72852 \pm 0.00031$		& $52 \pm 27$		& 1.9 	&this work\\
20080612					& 572 		& $ 2454629.81483 \pm 0.00083$		& $90\pm 71$		& 1.3  	&this work\\
20080727					& 609		& $ 2454674.65443 \pm 0.00085$		& $-5 \pm 73$		& -0.1 	&this work\\
20090428					& 836		& $ 2454949.75803 \pm 0.00065$		&$-21\pm 56$		& -0.4 	&this work\\
20090504					& 841		& $ 2454955.81712 \pm 0.00110$		&$-61\pm 92$		& -0.7 	&this work\\
20090510					& 846		& $ 2454961.87779 \pm 0.00069$		&$ 35 \pm 60$		&  0.6 	&this work\\
20090521					& 855		& $ 2454972.78553 \pm 0.00044$		&$ 81\pm 38$		&  2.2 	&this work\\
20090612					& 873		& $ 2454994.59691 \pm 0.00060$		&$-179 \pm 52$	& -3.4 	&this work\\
20090613					& 874		& $ 2454995.80961 \pm 0.00079$		&$-111 \pm 68$	& -1.6 	&this work\\
20090728					& 911		& $ 2455040.65158 \pm 0.00093$		&$ -1 \pm 80$		& -0.0 	&this work\\
\enddata
\tablenotetext{a} {~Time has been adjusted from the published times into the $BJD_{TT}$ time system by adding UTC-TT conversion of 64.184 sec.}
\tablenotetext{b} {~Midtime value when the $r'$ and $z'$ light curves are combined and fit independently.}
\tablenotetext{c} {~Light curve from \citet{Pont2007}; the midtime reported is from our independent fit of the original photometry.}
\tablenotetext{d} {~Midtime value when the $R$ and $V$ light curves are combined and fit independently.}
\label{table:ogle56ominusc}
\end{deluxetable}

\clearpage


\clearpage

\clearpage

\end{document}